\documentclass[draftclsnofoot, 12pt, onecolumn]{IEEEtran}

\usepackage{graphicx}
\usepackage{color}
\usepackage{placeins}
\usepackage{float}
\usepackage{tabularx,colortbl}
\usepackage{amsmath}
\usepackage{multirow}
\usepackage{hyperref}

\begin{document}
%
\title{All-Digital Self-interference Cancellation Technique for Full-duplex Systems}

\author{Elsayed Ahmed, and Ahmed M. Eltawil\footnote{Elsayed Ahmed and Ahmed M. Eltawil are with the Department of Electrical Engineering and Computer Science at the University of California, Irvine, CA, USA (e-mail: \{ahmede, aeltawil\}@uci.edu).}}

\maketitle


\begin{abstract}
Full-duplex systems are expected to double the spectral efficiency compared to conventional half-duplex systems if the self-interference signal can be significantly mitigated. Digital cancellation is one of the lowest complexity self-interference cancellation techniques in full-duplex systems. However, its mitigation capability is very limited, mainly due to transmitter and receiver circuit's impairments. In this paper, we propose a novel digital self-interference cancellation technique for full-duplex systems. The proposed technique is shown to significantly mitigate the self-interference signal as well as the associated transmitter and receiver impairments. In the proposed technique, an auxiliary receiver chain is used to obtain a digital-domain copy of the transmitted Radio Frequency (RF) self-interference signal. The self-interference copy is then used in the digital-domain to cancel out both the self-interference signal and the associated impairments. Furthermore, to alleviate the receiver phase noise effect, a common oscillator is shared between the auxiliary and ordinary receiver chains. A thorough analytical and numerical analysis for the effect of the transmitter and receiver impairments on the cancellation capability of the proposed technique is presented. Finally, the overall performance is numerically investigated showing that using the proposed technique, the self-interference signal could be mitigated to $\sim$3dB higher than the receiver noise floor, which results in up to 76\% rate improvement compared to conventional half-duplex systems at 20dBm transmit power values.
\end{abstract}

\begin{IEEEkeywords}
Full-duplex systems, digital self-interference cancellation, transceiver nonlinearities, phase noise.
\end{IEEEkeywords}

%
\IEEEpeerreviewmaketitle

\section{Introduction}
Recently, full-duplex transmission, where bidirectional communication is carried out over the same temporal and spectral resources, was introduced as a promising mechanism that could potentially double the spectral efficiency of wireless systems. The main limitation impacting full-duplex transmission is managing the strong self-interference signal imposed by the transmit antenna on the receive antenna within the same transceiver. For a full-duplex system to achieve its maximum efficiency, the self-interference signal has to be significantly suppressed to the receiver's noise floor. For instance, in WiFi systems, the transmit power can go up to 20dBm and the typical receiver noise floor is at $-$90dBm, which requires a total of 110dB self-interference cancellation for proper operation of a full-duplex system. In case the achieved amount of self-interference cancellation does not reach the receiver noise floor, the residual self-interference power will degrade the System's Signal to Noise Ratio (SNR) and thus negatively impact throughput.

Recently, several publications~\cite{Ref1}-\cite{Ref15} have considered the problem of self-interference cancellation in full-duplex systems by investigating different system architectures and self-interference cancellation techniques to mitigate the self-interference signal. Self-interference cancellation techniques are divided into two main categories: passive suppression, and active cancellation. In passive suppression~\cite{Ref10}-\cite{Ref15}, the self-interference signal is suppressed in the propagation domain before it is processed by the receiver circuitry. In active cancellation techniques~\cite{Ref5}-\cite{Ref9}, the self-interference signal is mitigated by subtracting a processed copy of the transmitted signal from the received signal. Active cancellation techniques could be divided into digital and analog cancellation techniques based on the signal domain (digital-domain or analog-domain) where the self-interference signal is subtracted.

Typical full-duplex systems deploy both passive suppression and active cancellation techniques to achieve significant self-interference mitigation~\cite{Ref13}-\cite{Ref15}. Since the cancellation process is performed in the digital-domain, digital self-interference cancellation techniques are the lowest complexity active cancellation techniques. However, the self-interference cancellation amount achieved by conventional digital cancellation techniques is very limited, mainly due to hardware imperfections. More specifically, the experimental and analytical results in~\cite{Ref16}-\cite{Ref18} show that transceiver phase noise and nonlinearities are the two main performance limiting factors in conventional digital cancellation techniques. 

In this paper, we propose a novel digital self-interference cancellation technique that eliminates all transmitter impairments, and significantly mitigates the receiver phase noise and nonlinearity effects. With the proposed technique and transceiver architecture, digital self-interference cancellation is no longer limited by the transceiver phase noise or nonlinearities.   

Recently several impairments suppression techniques have been proposed to alleviate the transceiver phase noise and nonlinearity effects on the self-interference cancellation performance. In~\cite{Ref7,Ref8}, modified digital cancellation techniques with nonlinearity suppression algorithms are proposed to mitigate the transceiver nonlinearities associated with the self-interference signal. In~\cite{Ref19} several phase noise estimation and suppression techniques for full-duplex systems are investigated showing that, even using highly complex techniques, only $\sim$2dB of phase noise suppression could be achieved. Accordingly, phase noise suppression remains a considerable issue in full-duplex systems. 

In addition to digital-domain impairments' suppression techniques, several full-duplex transceiver architectures are proposed to cancel out the impairments associated with the self-interference signal~\cite{Ref1,Ref7}. The main idea in such architectures is to obtain a copy of the transmitted Radio Frequency (RF) self-interference signal including all impairments and subtract it from the received signal in the RF domain. Since the obtained copy includes all transmitter impairments, the subtraction process is supposed to eliminate both the self-interference signal and the noise associated with it. In~\cite{Ref1}, a copy of the transmitted RF self-interference signal is passed through a variable attenuator and phase shifter then subtracted from the received signal in the RF domain. Since only one variable attenuator and phase shifter are used, these techniques will only mitigate the main component of the self-interference signal without mitigating the self-interference reflections. This issue has been handled in~\cite{Ref7}, where a multi-tap RF Finite Impulse Response (FIR) filter is used instead of the single attenuator. In this case, both main and reflected self-interference components (including the associated noise) are significantly mitigated at the receiver input. However, the size and power consumption of the RF FIR filter limits the applicability of such techniques.

In contrast with RF and analog cancellation techniques, we propose a novel all-digital self-interference cancellation technique based on a new full-duplex transceiver architecture that significantly mitigates transmitter and receiver impairments. In the proposed technique (shown in figure~\ref{Fig1Label}), an auxiliary receiver chain is used to obtain a digital-domain copy of the transmitted RF self-interference signal, which is then used to cancel out the self-interference signal and the associated transmitter impairments in the digital-domain. The auxiliary receiver chain has identical components as the ordinary receiver chain to emulate the effect of the ordinary receiver chain on the received signal. Furthermore, in order to alleviate the receiver phase noise effect, the auxiliary and ordinary receiver chains share a common oscillator. The proposed technique is shown to significantly mitigate the transmitter and receiver impairments without the necessity for highly complex RF cancellation techniques. The main advantage of the proposed technique is that all signal processing is performed in the digital-domain, which significantly reduces the implementation complexity.

In this paper, first, the performance of the proposed techniques is analytically and numerically investigated using a detailed full-duplex signal model that includes all major transmitter and receiver impairments. More specifically, transmitter and receiver nonlinearities, transmitter and receiver phase noise, receiver Gaussian noise, receiver quantization noise, and channel estimation errors are accounted for. Second, a thorough analytical and numerical analysis for the effect of each one of the transceiver impairments on the cancellation capability of the proposed technique is presented. Third, the performance and design tradeoffs involved with the proposed technique are also investigated. The analyses show that the proposed technique significantly mitigates the transceiver phase noise and nonlinearity effects, such that they are no longer the main performance limiting factors.

Finally, the overall full-duplex system performance using a combination of the proposed digital cancellation technique and the practical passive suppression techniques proposed in~\cite{Ref13,Ref14} is numerically investigated. The results show that, the proposed technique significantly mitigates the self-interference signal to $\sim$3dB higher than the receiver noise floor, which results in up to 67-76\% rate improvement compared to conventional half-duplex systems at 20dBm transmit power values.  

The rest of the paper is organized as follows: in section II, the detailed system model of the proposed architecture is presented. In section III, the analytical analysis of the proposed self-interference cancellation technique is presented. The overall full-duplex system performance is investigated in section IV. Section V presents the conclusion.

\section{System Model}
\begin{figure}[!ht]
\begin{center}
\noindent
  \includegraphics[width=6.5in]{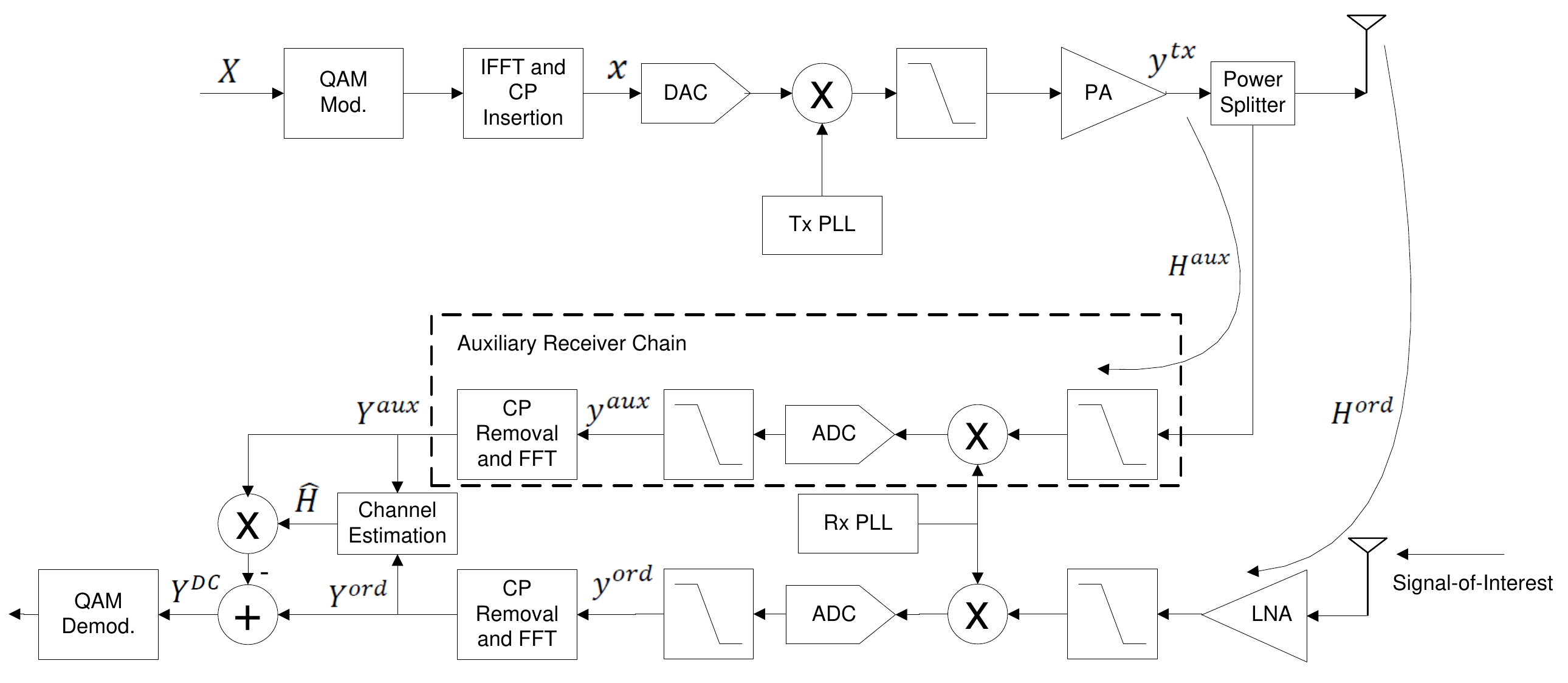}
  \caption{Detailed block diagram for the proposed digital self-interference cancellation technique.\label{Fig1Label}}
\end{center}
\end{figure}

In this section, the system model of the proposed full-duplex transceiver architecture is described in details. Figure~\ref{Fig1Label} shows a detailed block diagram for the proposed digital self-interference cancellation technique based on a new full-duplex transceiver architecture. The transceiver consists of the ordinary transmit and receive chains in addition to one auxiliary receiver chain used for self-interference cancellation. At the transmitter side, the information signal $X$ is Orthogonal Frequency Division Multiplexing (OFDM) modulated, then up-converted to the RF frequency. The up-converted signal is then filtered, amplified, and transmitted through the transmit antenna. A fraction of the amplified signal is fed-back as input to the auxiliary receiver chain. The signal power at the auxiliary receiver input is controlled through the power splitter; therefore, the Low Noise Amplifier (LNA) could be omitted from the auxiliary receiver chain. The feed-back and ordinary-received signals are down-converted to base-band through the auxiliary and ordinary receiver chains respectively. The auxiliary and ordinary receiver chains are identical and share the same Phase Locked Loop (PLL). The channel transfer function $H^{aux}$ represents the wired channel from the Power Amplifier (PA) output to the auxiliary receiver. While the channel transfer function $H^{ord}$ represents the self-interference wireless channel.

The output of the auxiliary and the ordinary receiver chains are fed to a channel estimation block to obtain an estimate for the ratio between the ordinary and auxiliary channels ($H^{ord}/H^{aux}$). The channel estimation process is performed using time-orthogonal training sequences transmitted at the beginning of each data frame. The estimated channel is then multiplied with the auxiliary receiver output, and the multiplication output is subtracted from the received signal to obtain an interference-free signal.

In this paper, the main transmitter and receiver impairments are considered. More specifically, transmitter and receiver phase noise, transmitter and receiver nonlinearities, Analog-to-Digital Converter (ADC) quantization noise, and receiver Gaussian noise. Since the feed-back signal is obtained from the PA output which contains a copy of the transmitter impairments, the proposed architecture can significantly mitigate all transmitter impairments. In addition, the receiver phase noise effect is mitigated by means of sharing the same PLL between the auxiliary and ordinary receiver chains. In order to investigate the performance of the proposed technique, we first present a signal model with a detailed modeling for each one of the transceiver impairments. 

In the presence of the transmitter phase noise $\phi^{tx}$ and the PA nonlinear distortion signal $d^{tx}$, the transmitted signal at the PA output can be written as
\begin{equation}\label{eq:1}
y^{tx}(t)=x(t)e^{j\left(2\pi f_c t+\phi^{tx}(t)\right)}+d^{tx}(t)\text{,}
\end{equation}
where $x$ is the transmitter base-band signal, $d^{tx}$ is the transmitter nonlinear distortion due to the PA, and $f_c$ is the carrier frequency. At the auxiliary receiver output, the digital base-band signal $y^{aux}$ can be written as
\begin{equation}\label{eq:2}
y_n^{aux}=\left(y_n^{tx}*h_n^{aux}\right)e^{j\phi^{rx}_n}+q_n^{aux}+z_n^{aux}\text{,}
\end{equation}
where $*$ denotes convolution process, $n$ is the sample index, $y^{tx}_n$ is the digital base-band representation of $y^{tx}(t)$, $h^{aux}$ is the wired channel from the PA output to the auxiliary receiver input, $\phi^{rx}$ is the receiver phase noise process, $q^{aux}$ is the auxiliary receiver ADC quantization noise, and $z^{aux}$ is the auxiliary receiver Gaussian noise. Similarly, the digital base-band signal $y^{ord}$ at the ordinary receiver output can be written as
\begin{equation}\label{eq:3}
y_n^{ord}=\left(y_n^{tx}*h_n^{ord}\right)e^{j\phi^{rx}_n}+d_n^{rx}+q_n^{ord}+z_n^{ord}+ s_n^{soi}\text{,}
\end{equation}
where superscript $ord$ refers to the ordinary receiver chain signals, $d^{rx}$ is the receiver nonlinear distortion due to the LNA, and $s^{soi}$ is the received signal-of-interest (including the signal-of-interest channel and all impairments). After digial self-interference cancellation, the interference-free signal $Y^{DC}$ can be written as 
\begin{equation}\label{eq:4}
Y_k^{DC}=Y_k^{ord}-Y_k^{aux} \hat{H}_k\text{,}
\end{equation}
where uppercase letters denote the frequency domain representation of the corresponding signals, $k$ is the subcarrier index, and $\hat{H}$ is an estimate for the ratio between the ordinary and auxiliary channels ($H^{ord}/H^{aux}$) calculated in the Least Square (LS) form as
\begin{equation}\label{eq:5}
\hat{H}_k=\frac{Y_k^{ord}}{Y_k^{aux}}\text{.}
\end{equation}
For a complete signal model, a detailed description for the impairments and channel modeling is presented in the following subsections.
\subsection{Transceiver Nonlinearities}
In practical systems, the main sources of the system nonlinearity are the power amplifier at the transmitter side, and the LNA at the receiver side. Generally, for any nonlinear block, the output signal $y$ can be written as a polynomial function of the input signal $g$ as follows~\cite{Ref20} 
\begin{equation}\label{eq:6}
y(t)=\sum_{m=1}^{M}\alpha_m g(t)^m\text{,}
\end{equation}
where the first term represents the linear component, and higher order terms contribute to the nonlinear distortion. With simple analysis, it can be easily shown that only the odd orders of the polynomial contribute to the in-band distortion~\cite{Ref20}. Accordingly, equation~\eqref{eq:6} can be further simplified and written is the digital base-band domain as 
\begin{equation}\label{eq:7}
y_n=\sum_{m=1,m \ is \ odd}^{M} \alpha_m g_n \left|g_n\right|^{m-1}\text{,}
\end{equation}
where $g_n$ and $y_n$ are the digital base-band representation of the input and output of the nonlinear block. For the PA nonlinearity, the digital base-band representation of the input signal is $g_n = x_n e^{j\phi^{tx}_n}$. While, for the LNA nonlinearity, the LNA input is the transmitted signal after going through the wireless channel, i.e. $g_n=(y_n^{tx}*h^{ord}_n)$.   

Typically, a limited number of orders contribute to the major distortion and higher orders could be neglected. In practical systems, the nonlinearity is typically characterized by the third-order intercept point (IP3), which is defined as the point at which the power of the third harmonic is equal to the power of the first harmonic~\cite{Ref21}. While the signal model is valid for any nonlinearity order, the numerical analysis in this paper is limited to the third-order nonlinearity.
\subsection{Transceiver Phase Noise}
Generally, modeling the phase noise process depends on the oscillator type. There are two main oscillator types: free-running oscillators and PLL based oscillators. In free-running oscillators the phase noise could be modeled as a Wiener process~\cite{Ref22} where the phase error at the $n^{th}$ sample is related to the previous one as $\phi_n = \phi_{n-1} + \beta$, where $\beta$ is a Gaussian random variable with zero mean and variance $\sigma^2=4\pi^2f_c^2CT_s$. In this notation $T_s$ describes the sample interval and $C$ is an oscillator dependent parameter that determines its quality. The oscillator parameter $C$ is related to the 3dB bandwidth $f_{3dB}$ of the phase noise Lorentzian spectrum by $C=f_{3dB}/\pi f_c^2$. 

\begin{figure}[!ht]
\begin{center}
\noindent
  \includegraphics[width=2.5in]{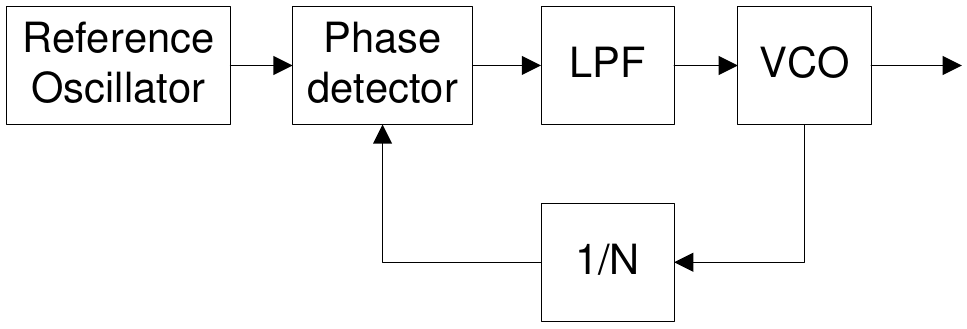}
  \caption{Block diagram for a PLL based oscillator.\label{Fig2Label}}
\end{center}
\end{figure}

In PLL based oscillators, as shown in figure~\ref{Fig2Label}, the voltage controlled oscillator (VCO) output is controlled through a feed-back loop that involves a phase detector and low-pass filter (LPF). The purpose of the feed-back loop is to lock the phase of the VCO output with the phase of a high quality reference oscillator. As shown in~\cite{Ref23}, the PLL output phase noise can be modeled as Ornstein-Uhlenbeck process with auto-correlation function calculated as
\begin{equation}\label{eq:8}
E\left[e^{j\Delta\phi_{mn}}\right] = e^{\frac{-4\pi^2 f_c^2}{2}\left(C T_s |m-n| + 2\sum_{i=0}^{n_0}\left(\mu_i+v_i\right)\left(1-e^{-\lambda_i T_s |m-n|}\right)\right)}\text{,}
\end{equation}
where ($n_0$, $\mu$ , $v$, $\lambda$) are PLL specific parameters that are function of the PLL loop filter design\footnote{See~\cite{Ref23} for detailed description on how these parameters are calculated.}. The Power Spectral Density (PSD) of the PLL phase noise is obtained by taking the Fourier transform of equation~\eqref{eq:8}. Since PLL-based oscillators are commonly used in wireless systems, the numerical analysis in this paper is performed using PLL-based oscillators.
\subsection{Gaussian and Quantization Noise }
The receiver Gaussian noise $z_n$ represents the additive noise inherent in the receiver circuits, and is usually specified by the receiver noise figure. The receiver Gaussian noise is modeled as a Gaussian random variable with zero mean and variance $\sigma^2_z = P_{th}N_f$, where $P_{th}$ is the thermal noise power in a 50ohm source resistance, and $N_f$ is the overall receiver noise figure. Generally, the overall receiver noise figure is directly proportional to the input signal level. The overall receiver noise figure can be written as a function of the noise figure of each individual block as~\cite{Ref21}
\begin{equation}\label{eq:88}
N_f=N_{LNA}+\sum_{l=1}^{L}\frac{N_l-1}{\alpha^2}\text{,}
\end{equation}
where $N_{LNA}$ is the LNA noise figure, $N_l$ is the noise figure of the $l^{th}$ block after the LNA block, and $\alpha^2$ is the LNA power gain. Equation~\eqref{eq:88} assumes that all blocks following the LNA have unity gain.

The ADC quantization noise $q_n$ is a uniformly distributed noise introduced by the ADC due to the signal quantization. For an $m$ bits ADC, the total ADC quantization noise power is calculated as~\cite{Ref24}
\begin{equation}\label{eq:9}
P_q=\frac{1}{\alpha^2} \frac{1}{12\cdot 2^{2m-2}} =\frac{\sigma_q^2}{\alpha^2}\text{,}
\end{equation}
where $\sigma_q^2=1/12\cdot 2^{2m-2}$ is the quantization noise variance. It has to be noted that, although the auxiliary and the ordinary receiver chains have identical components, the Gaussian and quantization noises are independent for the two receiver chains.
\subsection{Wireless Channel Modeling}
Generally, in full-duplex systems, the self-interference channel consists of two main components: The Line-of-Sight (LOS) component due to the direct link between the transmit and receive antennas, and the non-LOS component due to the signal reflections. Accordingly, the first tap of the self-interference channel could be modeled as Rician fading with Rician factor $k$ and the remaining channel taps are modeled as Rayleigh fading with variance $k$. The Rician factor $k$ represents the power ratio between the LOS and the reflective components of the channel. The experimental characterization presented in~\cite{Ref13} show that for typical indoor environments with antenna separation of 35-50cm, the self-interference channel Rician factor is approximately 20-25dB. 

The experimental results in~\cite{Ref13} also show that the self-interference channel Rician factor is inversely proportional to the achieved passive suppression amount. For instance, using omni-directional antenna with antenna separation of 50cm could achieve up to 28dB passive suppression, and the self-interference channel Rician factor in this case is $\sim$25dB. On the hand, using directional antennas could achieve up to 45dB of passive suppression, however, the self-interference channel Rician factor decreases to $\sim$0dB. The main reason for this inverse relation is that passive suppression techniques tends to significantly mitigate the LOS component and slightly mitigates the reflections.  

In this paper, the analyses are based on the two-antenna transceiver architecture shown in figure 1. However, another architecture where the transmit and receive antennas are replaced by a circulator and single antenna could be used~\cite{Ref7}. In this case the achieved passive suppression will be limited by the circulator coupling (typically 20dB). On the other hand, the self-interference channel for the single-antenna architecture is expected to have larger Rician factor, mainly due to the fact that the reflections have to make a roundtrip in order to be reflected back to the same antenna. The impact of the passive suppression amount and channel Rician factor on the cancellation performance is discussed in Section III.

\section{Self-interference Cancellation Analysis}
The main idea of the proposed cancellation technique is to obtain a copy of the transmitted self-interference signal including all transmitter impairments, and use this copy for digital-domain self-interference cancellation at the receiver side. Hypothetically speaking, if both auxiliary and ordinary receiver chains are impairment-free, the proposed architecture should be able to totally eliminate both the self-interference signal and the transmitter impairments. However, due to the receiver impairments and the channel estimation errors, perfect self-interference cancellation is not possible. In fact, receiver impairments and channel estimation errors introduce certain limitations on the self-interference cancellation capability. In order to understand these limitations, we analytically and numerically investigate the impact of the receiver impairments and channel estimation errors on the self-interference cancellation capability of the proposed technique.

For a clear understanding of the impairments effect and the involved tradeoffs, each impairment is analyzed individually (i.e. the system is analyzed in the presence of one receiver impairment at a time). At the end, the overall performance in the presence of all impairments is investigated. In each analysis, all transmitter impairments are considered; only the receiver impairments are considered individually. During the analysis of the individual impairments, the auxiliary and ordinary channel transfer functions ($H^{aux}, H^{ord}$) are assumed to be perfectly known. The effect of channel estimation errors is studied in a separate subsection. The numerical analyses are based on a 20MHz OFDM-based system with 64 subcarriers per OFDM symbols as in IEEE802.11 systems~\cite{Ref27}. The carrier frequency is set to 2.4GHz.  
\subsection{Impact of Gaussian and Quantization Noise}
In the presence of only Gaussian and quantization noise, the auxiliary and ordinary receiver outputs (Equation~\eqref{eq:2}~and~\eqref{eq:3}) can be rewritten as
\begin{equation}\label{eq:10}
y_n^{aux}=y_n^{tx}*h_n^{aux}+q_n^{aux}+z_n^{aux}\text{,}
\end{equation}
\begin{equation}\label{eq:11}
y_n^{ord}=y_n^{tx}*h_n^{ord}+q_n^{ord}+z_n^{ord}+s_n^{soi}\text{,}
\end{equation}
After self-interference cancellation, the frequency domain representation of the interference-free signal $Y^{DC}$ can be written as
\begin{equation}\label{eq:12}
Y_k^{DC}=Y_k^{ord}-\frac{H_k^{ord}}{H_k^{aux}} Y_k^{aux} =S_k^{soi}+Q_k^{ord}+Z_k^{ord}-\bar{Q}_k^{aux}-\bar{Z}_k^{aux}\text{,}
\end{equation}
where $\bar{Q}^{aux}$ and $\bar{Z}^{aux}$ are the modified auxiliary quantization and Gaussian noise after multiplication with the channel transfer function $H^{ord}/H^{aux}$. 

Equation~\eqref{eq:12} sets the first performance limiting factor that results due to the Gaussian and quantization noise. In terms of power levels, since the noise terms are not correlated, the total noise power is calculated as the summation of the power of the four noise terms in~\eqref{eq:12}. As shown in~\eqref{eq:9} the quantization noise power is inversely proportional to the LNA power gain and number of ADC bits. Therefore, one can easily get rid of the quantization noise limitation by increasing the number of ADC bits. For instance, using 14-bits ADC will result is quantization noise power of $\sim-$90dBm at 0dB LNA gain. However, increasing the number of ADC bits slightly increases the hardware complexity. 

Typically, Gaussian noise dominates quantization noise, especially at high input power levels. Equation~\eqref{eq:88} shows that the Gaussian noise power is a function of the input signal power level, and the noise figure of individual receiver components. At low input signal power levels (i.e. high LNA gain) the Gaussian noise power is dominated by the LNA noise figure which is designed to be relatively small. However, at high input signal power levels, the Gaussian noise power is dominated by the noise figure of the components following the LNA, which typically have relatively high noise figure. As a practical example, the NI5791 transceiver datasheet~\cite{Ref25} show that the receiver Gaussian noise power is $-$163, and $-$145dBm/Hz at signal power levels of $-$25, and $-$5dBm respectively. This is equivalent to a total noise power of $-$90 and $-$72dBm in a 20MHz bandwidth. Accordingly, decreasing the Gaussian noise effect requires low input signal power levels, which could be achieved through good passive self-interference suppression, or by decreasing the transmit power.

\begin{figure}[!ht]
\begin{center}
\noindent
  \includegraphics[width=3.5in]{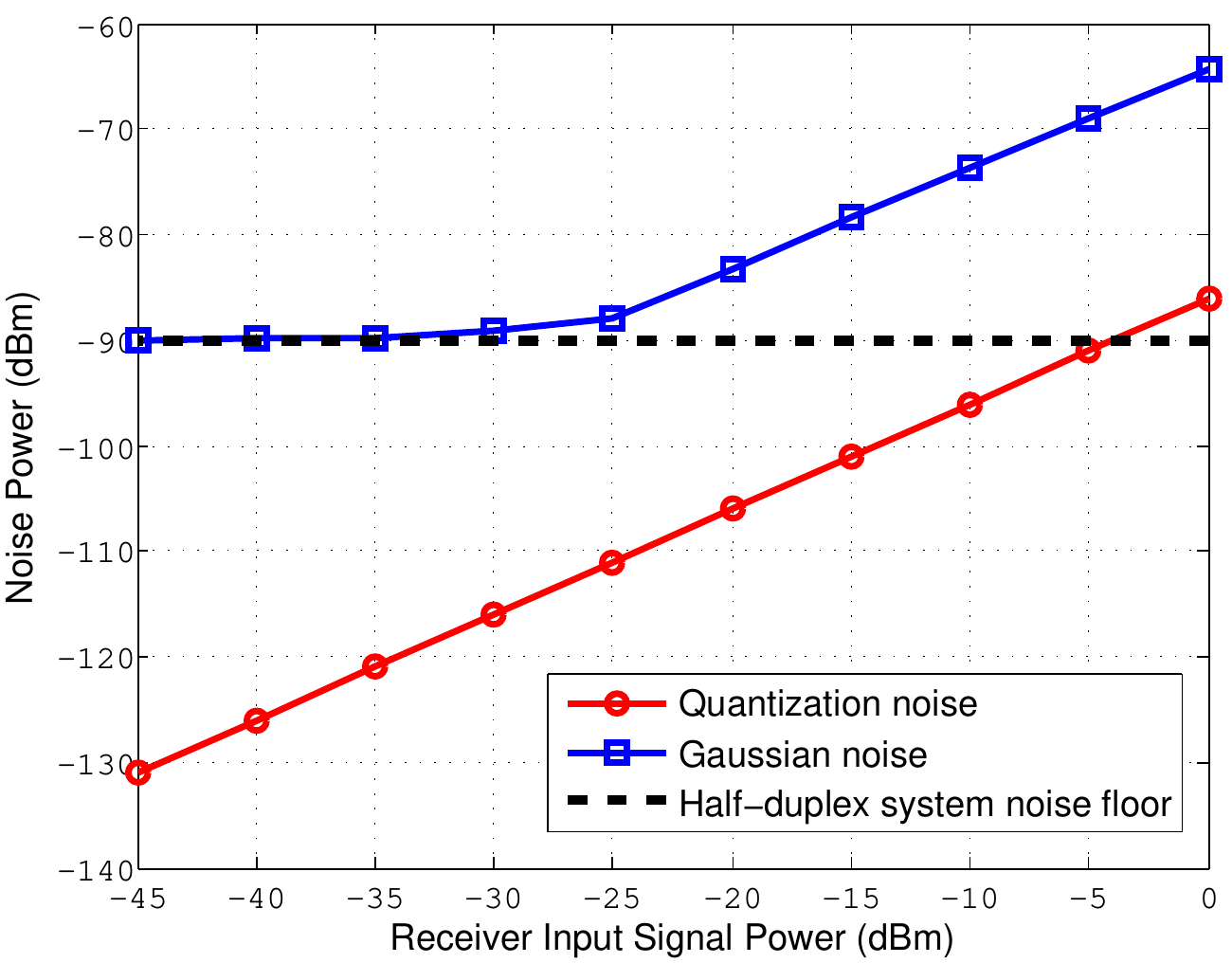}
  \caption{Quantization and Gaussian noise power at different receiver input signal power levels.\label{Fig3Label}}
\end{center}
\end{figure}

For more clarification, a numerical analysis is performed to investigate the impact of the Gaussian and quantization noise on the cancellation performance at different scenarios. In this analysis, the system is simulated using practical receiver parameters from the NI5791 transceiver datasheet~\cite{Ref25}. More specifically, the number of ADC bits is set to 14bits, the receiver Gaussian noise is $-$90dBm for LNA gains $\geq$25dB, and $-$72dBm at 5dB LNA gain. Figure~\ref{Fig3Label} shows the quantization and Gaussian noise power at different receiver input signal power levels. The half-duplex system noise floor is shown as a comparison reference. In half-duplex systems, since the input signal power is typically small, the noise floor is dominated by the LNA noise figure ($-$90dBm is this example). The results show that for input power levels $\leq-$30dBm, the proposed full-duplex system has the same noise floor as compared to half-duplex systems. However, at high received signal power levels, the full-duplex system's noise floor linearly increases with the received signal power, mainly due to the decrease of the LNA gain, which increases the overall receiver noise figure as shown in~\eqref{eq:88}.

Figure 3 show that at high received signal power levels, Gaussian noise is a considerable performance limiting factor. The Gaussian noise effect could be reduced by using either good passive suppression techniques as in~\cite{Ref13,Ref14}, or using the Balun RF cancellation proposed in~\cite{Ref1}. In the Balun cancellation technique, part of the self-interference signal is mitigated before the signal goes through the LNA, which reduces the overall receiver noise figure and thus the Gaussian noise power.    
\subsection{Impact of Receiver Phase Noise}
In this analysis, the receiver is assumed to have only the phase noise impairment. Furthermore, the auxiliary and ordinary receiver chains are sharing the same PLL, and thus have the same phase noise signal. In the presence of only receiver phase noise, the auxiliary and ordinary receiver outputs (Equation~\eqref{eq:2}~and~\eqref{eq:3}) can be rewritten as
\begin{equation}\label{eq:13}
y_n^{aux}=\left(y_n^{tx}*h_n^{aux}\right)e^{j\phi^{rx}_n}\text{,}
\end{equation}
\begin{equation}\label{eq:14}
y_n^{ord}=\left(y_n^{tx}*h_n^{ord}\right)e^{j\phi^{rx}_n} + s_n^{soi}\text{,}
\end{equation}
By performing Discrete Fourier Transform (DFT), the frequency domain representation of~\eqref{eq:13}~and~\eqref{eq:14} can be written as
\begin{equation}\label{eq:15}
Y_k^{aux}=\sum_{l=0}^{N-1}Y_l^{tx} H_l^{aux} J_{k-l}^{rx}\text{,}
\end{equation}
\begin{equation}\label{eq:16}
Y_k^{ord}=\sum_{l=0}^{N-1}Y_l^{tx} H_l^{ord} J_{k-l}^{rx} + S_k^{soi}\text{,}
\end{equation}
where $N$ is the number of subcarriers per OFDM symbol, and $J^{rx}$ is the DFT coefficients of the phase noise process $e^{j\phi^{rx}}$ calculated as
\begin{equation}\label{eq:17}
J_k^{rx}=\sum_{n=0}^{N-1}e^{j\phi_n^{rx}} e^{-j2\pi \frac{nk}{N}}\text{.}
\end{equation}
Since $H^{aux}$ is a wired channel, it can be assumed that $H^{aux}$ has a frequency flat response. Accordingly, equation~\eqref{eq:15} can be simplified as
\begin{equation}\label{eq:18}
Y_k^{aux}=H_k^{aux} \sum_{l=0}^{N-1}Y_l^{tx}J_{k-l}^{rx}\text{.}
\end{equation}
After self-interference cancellation, the interference-free signal $Y^{DC}$ can be written as
\begin{eqnarray}\label{eq:19}
Y_k^{DC}&=& Y_k^{ord}-\frac{H_k^{ord}}{H_k^{aux}} Y_k^{aux} =S_k^{soi}+\sum_{l=0}^{N-1}Y_l^{tx} H_l^{ord} J_{k-l}^{rx}- H_k^{ord}\sum_{l=0}^{N-1}Y_l^{tx} J_{k-l}^{rx} \nonumber \\
&=& S_k^{soi}+\sum_{l=0,l\neq k}^{N-1}\left(H_l^{ord}-H_k^{ord}\right)Y_l^{tx} J_{k-l}^{rx}\text{.}
\end{eqnarray}
The second term in the right hand side of~\eqref{eq:19} represents the residual self-interference signal due to the receiver phase noise effect. 

According to~\eqref{eq:19}, the residual self-interference (RSI) power $P_{RSI}$ can be calculated as
\begin{eqnarray}\label{eq:20}
P_{RSI} &=& E\left[\left|\sum_{l=0,l\neq k}^{N-1}\left(H_l^{ord}-H_k^{ord}\right)Y_l^{tx}J_{k-l}^{rx} \right|^2 \right] \nonumber \\
&=& \sum_{l=0,l\neq k}^{N-1}E\left[\left|\left(H_l^{ord}-H_k^{ord}\right)\right|^2 \right] E\left[\left|Y_l^{tx}\right|^2\right]E\left[\left|J_{k-l}^{rx}\right|^2\right] \nonumber \\
&=& P^{tx} \sum_{l=0,l\neq k}^{N-1}E\left[\left|\left(H_l^{ord}-H_k^{ord}\right)\right|^2\right] E\left[\left|J_{k-l}^{rx}\right|^2\right]\text{,}
\end{eqnarray}
where $E\left[\left|Y^{tx}\right|^2\right]=P^{tx}$ is the transmit power. Decomposing $H^{ord}$ into LOS and non-LOS components where $H^{ord}_k = H^{ord,los} + H^{ord,nlos}_k$ equation~\eqref{eq:20} can be further simplified as   
\begin{equation}\label{eq:21}
P_{RSI}=P^{tx} \sum_{l=0,l\neq k}^{N-1} E\left[\left|\left(H_l^{ord,nlos}-H_k^{ord,nlos}\right)\right|^2\right] E\left[\left|J_{k-l}^{rx}\right|^2\right]\text{.}
\end{equation}
An upper bound for the residual self-interference power is obtained when $H^{ord,nlos}_ k$ and $ H^{ord,nlos}_l$ are uncorrelated for all $k\neq l$. In this case the upper bound of the residual self-interference power in~\eqref{eq:21} is calculated as  
\begin{equation}\label{eq:22}
P_{RSI}^U=P^{tx} \sum_{l=0,l\neq k}^{N-1} \left(E\left[\left|H_l^{ord,nlos}\right|^2\right] + E\left[\left|H_k^{ord,nlos}\right|^2\right]\right) E\left[\left|J_{k-l}^{rx}\right|^2\right]
=2P^{tx} P^{nlos} P^{PN}\text{.}
\end{equation}
where $P^{nlos}$ is the power of the non-LOS component of the self-interference channel, $P^{PN}$ is the total in-band phase noise power, and the factor of 2 is due to the subtraction of uncorrelated random variables. 
On the other hand, it is obvious that the lower bound of the residual self-interference power is zero. The lower bound is achieved when the self-interference channel has frequency-flat transfer function. Between the upper and lower bounds, the residual self-interference power is determined by the self-interference channel characteristics, mainly the correlation between the channel frequency response at different subcarriers. 

Typically, the correlation between the channel frequency response at different subcarriers is determined by the coherence bandwidth (CBW) of the channel. By definition, the CBW is defined as the frequency band in which the channel could be considered frequency flat. According to this definition, equation~\eqref{eq:21} can be rewritten as          
\begin{equation}\label{eq:23}
P_{RSI}=P^{tx}\sum_{l=0,k-l>|CBW|}^{N-1}E\left[\left|\left(H_l^{ord,nlos}-H_k^{ord,nlos}\right)\right|^2\right] E\left[\left|J_{k-l}^{rx}\right|^2\right]\text{,}
\end{equation}
where $H^{ord,nlos}_l - H^{ord,nlos}_k = 0$ within the CBW of the channel. Furthermore, as shown in~\cite{Ref19}, the phase noise process has a decaying power spectral density, such that the majority of the phase noise power is located around the DC carrier (i.e. $J_0$). Accordingly, the remaining terms where $k-l>|CBW|$ will be weighted by small phase noise values $\left|J_{k-l}\right|^2$.

Equation~\eqref{eq:23} shows that the proposed digital cancellation technique totally eliminates the phase noise associated with the LOS component of the channel in addition to part of the phase noise associated with the non-LOS component depending on the CBW. This is considered a significant improvement compared to conventional digital cancellation techniques where no phase noise elimination is achieved.  For example, if the self-interference channel has a Rician factor of 20dB (i.e. the non-LOS component is 20dB lower than the LOS component), the proposed technique will mitigate the phase noise by at least 20dB more than the conventional digital cancellation techniques. In addition, more mitigation will be achieved according to the CBW of the channel.   

\begin{figure}[!ht]
\begin{center}
\noindent
  \includegraphics[width=3.5in]{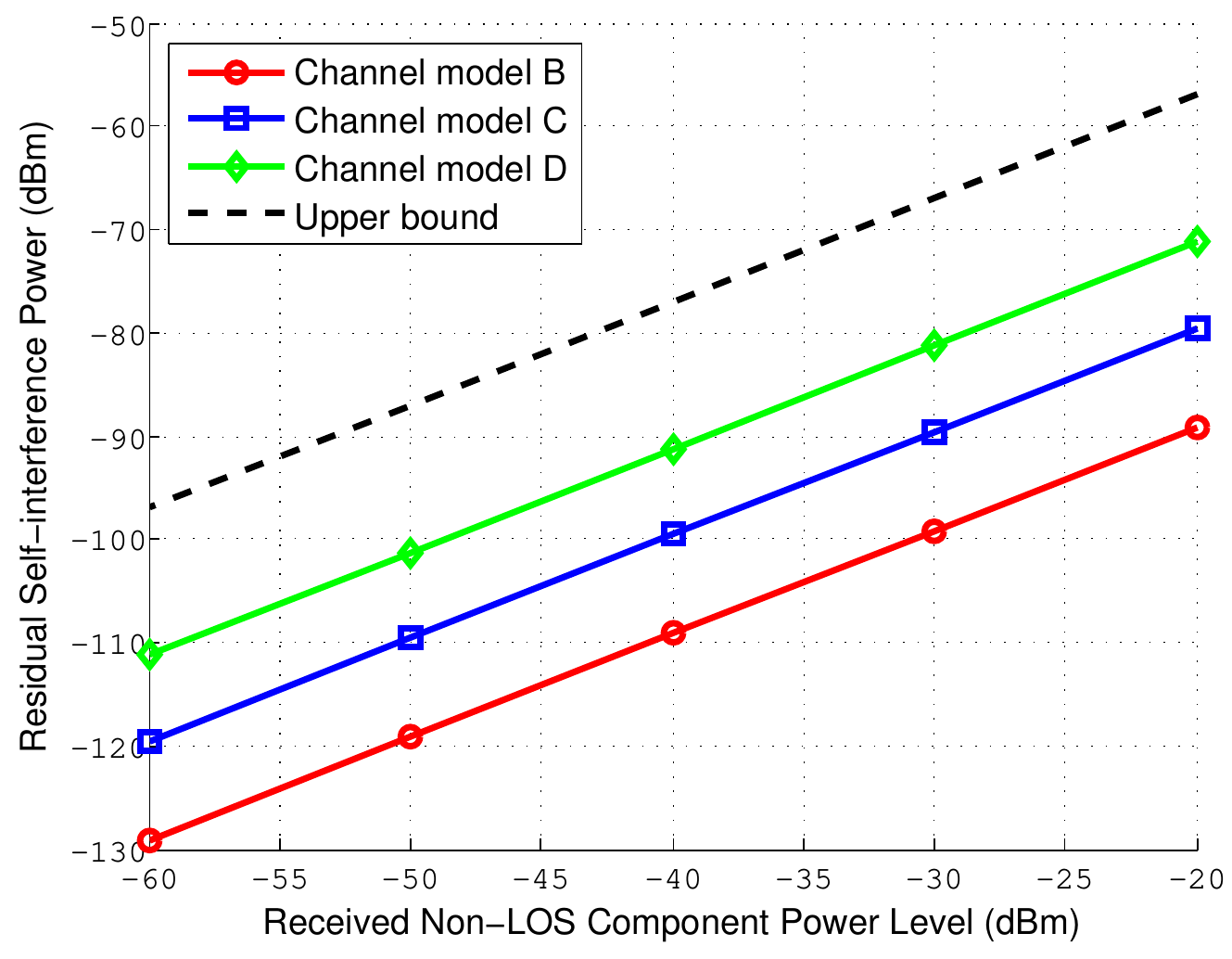}
  \caption{Residual self-interference power due to receiver phase noise effect for different channel models with $-$40dBc total in-band phase noise power.\label{Fig4Label}}
\end{center}
\end{figure}

In order to quantify the phase noise mitigation gain achieved due to the CBW of the channel, the system is simulated under three different channel models with different maximum delay spread (i.e. different CBW). The simulated channel models are the 'B', 'C', and 'D' TGn channel models defined in~\cite{Ref26}. The 'B', 'C', and 'D' channel models have a maximum delay spread of 80, 200, and 390ns respectively. Figure~\ref{Fig4Label} shows the residual self-interference power for the three channel models at different non-LOS component power levels, with total phase noise power of -40dBc. The results show that, larger CBW results in lower residual self-interference power (i.e. more phase noise mitigation). In terms of numbers, approximately 15-30dB reduction in residual self-interference power compared to the upper bound is achieved due to the correlation between the channel frequency responses within the CBW of the channel.  
\subsection{Impact of Channel Estimation Errors }
In the previous analyses, the channel transfer function $H^{ord}/H^{aux}$ was assumed to be perfectly known. However, in practical systems, the channel transfer function is obtained through a channel estimation process that results in channel estimation errors. In this section, the effect of the channel estimation errors on the cancellation performance is investigated. 

In practical indoor applications~\cite{Ref27}, the channel is estimated using training symbols transmitted at the beginning of each data frame. During the training interval, the Least Square (LS) estimator is used to obtain an estimate for the channel transfer function as 
\begin{equation}\label{eq:24}
\hat{H}_k=\frac{Y_k^{ord}}{Y_k^{aux}}\text{.}
\end{equation}
Since LS estimator is known to have estimation errors that are directly proportional to the noise level, channel estimation error effect should be analyzed in the presence of all system impairments. 

In the presence of receiver Gaussian, quantization, and phase noise, the frequency domain representation of the ordinary and auxiliary receiver outputs $Y^{ord}, Y^{aux}$ during the training interval can be written as \footnote{Note that the signal-of-interest is not transmitted during the self-interference training interval}
\begin{eqnarray}\label{eq:25}
Y_k^{ord} &=& \sum_{l=0}^{N-1}Y_l^{tx} H_l^{ord}J_{k-l}^{rx}+Q_k^{ord}+Z_k^{ord} \nonumber \\
&=& H_k^{ord}\sum_{l=0}^{N-1}Y_l^{tx} J_{k-l}^{rx} +\sum_{l=0}^{N-1}Y_l^{tx} \left(H_l^{ord}-H_k^{ord}\right)J_{k-l}^{rx} + Q_k^{ord}+Z_k^{ord} \nonumber \\
&=& H_k^{ord} \sum_{l=0}^{N-1}Y_l^{tx} J_{k-l}^{rx} + \eta_k\text{,}
\end{eqnarray}
\begin{eqnarray}\label{eq:26}
Y_k^{aux} &=& \sum_{l=0}^{N-1}Y_l^{tx} H_l^{aux}J_{k-l}^{rx}+Q_k^{aux}+Z_k^{aux} \ \ \ \ \ \ \ \ \ \ \ \ \ \ \ \ \ \ \ \ \ \ \ \ \ \ \ \ \ \ \ \ \ \ \ \nonumber \\
&=& H_k^{aux}\sum_{l=0}^{N-1}Y_l^{tx} J_{k-l}^{rx} + Q_k^{aux}+Z_k^{aux} \nonumber \\
&=& H_k^{aux} \sum_{l=0}^{N-1}Y_l^{tx} J_{k-l}^{rx} + \zeta_k\text{,}
\end{eqnarray}
where $\eta$, and $\zeta$ are the composite noise component in the ordinary and auxiliary receiver chains calculated as
\begin{equation}\label{eq:255}
\eta_k=\sum_{l=0}^{N-1}Y_l^{tx} \left(H_l^{ord}-H_k^{ord}\right)J_{k-l}^{rx} + Q_k^{ord}+Z_k^{ord}\text{,}
\end{equation}
\begin{equation}\label{eq:266}
\zeta_k=Q_k^{aux}+Z_k^{aux} \text{.}
\end{equation}
The first noise component in~\eqref{eq:255} represents the residual self-interference due to the receiver phase noise effect as described in~\eqref{eq:19}. Dividing~\eqref{eq:25} by~\eqref{eq:26} to obtain $\hat{H}$ we get
\begin{equation}\label{eq:27}
\hat{H}_k=\frac{H_k^{ord}\sum_{l=0}^{N-1}Y_l^{tx}J_{k-l}^{rx}+\eta_k}{H_k^{aux}\sum_{l=0}^{N-1}Y_l^{tx} J_{k-l}^{rx}+\zeta_k}\text{.}
\end{equation}
Since $\zeta_k \ll H^{aux}_k \sum_{l=0}^{N-1}Y^{tx}_k J^{rx}_{k-l}$, using the approximation of $(1+x)^{-1}\simeq 1-x, x\ll 1$, equation~\eqref{eq:27} can be approximated as
\begin{equation}\label{eq:28}
\hat{H}_k=\frac{H_k^{ord}}{H_k^{aux}}+\frac{\eta_k-\zeta_k}{H_k^{aux}\sum_{l=0}^{N-1}Y_l^{tx}J_{k-l}^{rx}} = \frac{H_k^{ord}}{H_k^{aux}}+ E_H\text{,}
\end{equation}
where $E_H$ is the channel estimation error due to the receiver impairments. Equation~\eqref{eq:28} shows that the channel estimation error is directly related to the summation of all noise components in both the auxiliary and ordinary receiver chains. One simple way to improve the channel estimation quality is by averaging the estimated channel $\hat{H}$ over multiple OFDM symbols as follows
\begin{equation}\label{eq:29}
\hat{H}_k=\frac{1}{M}\sum_{m=1}^M\frac{Y_{m,k}^{ord}}{Y_{m,k}^{aux}}\text{.}
\end{equation}
In this case, since the noise at different OFDM symbols are not correlated, the channel estimation error will be reduced by a factor of $M$, where $M$ is the number of training OFDM symbols. On the other hand, increasing the number of training symbols will negatively impact the overall system capacity. 

\begin{figure}[!ht]
\begin{center}
\noindent
  \includegraphics[width=6.5in]{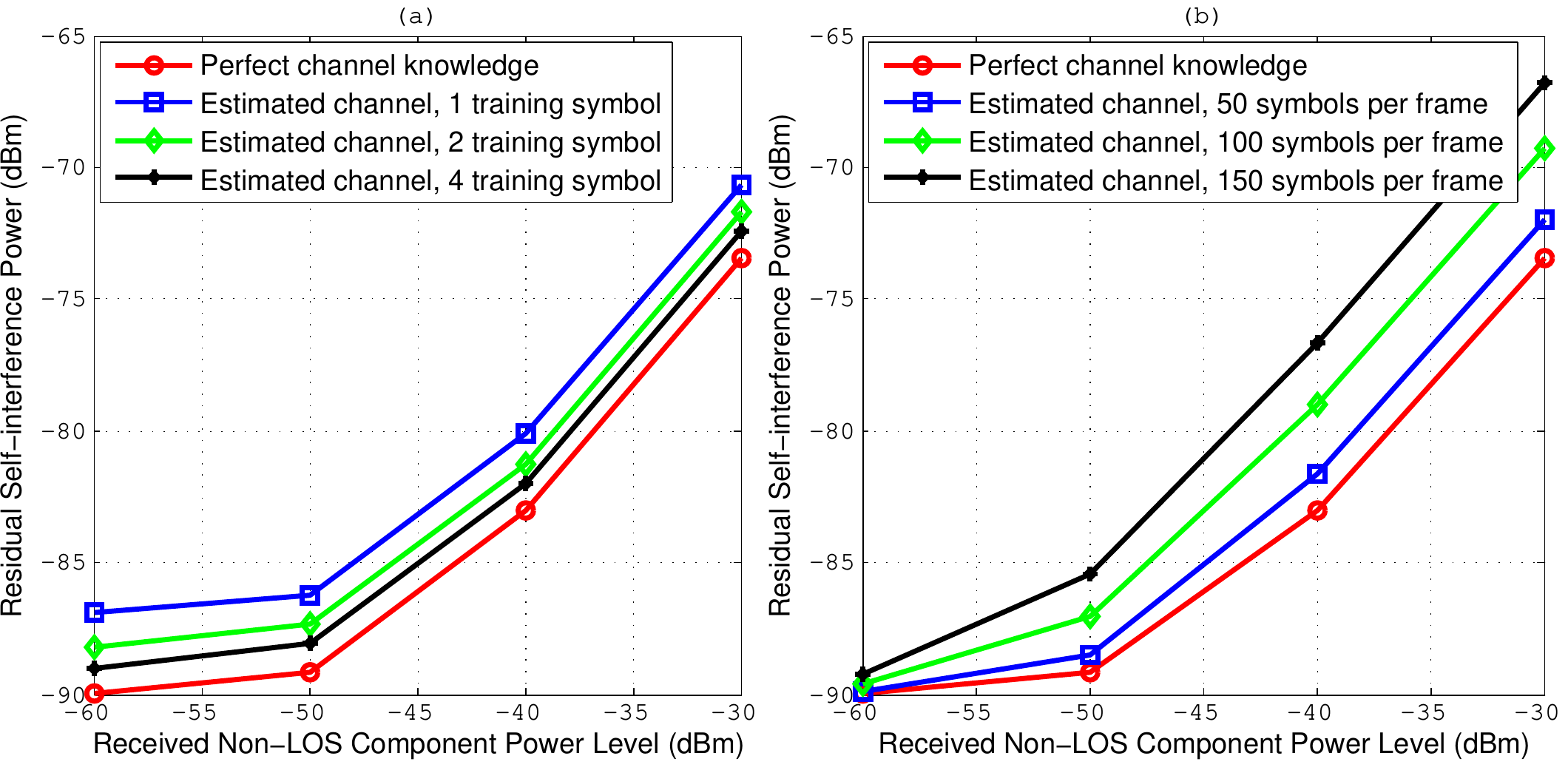}
  \caption{Residual self-interference power due to channel estimation error effect: (a) channel estimation error due to receiver noise, (b) channel estimation error due to fading effect.\label{Fig5Label}}
\end{center}
\end{figure}

After self-interference cancellation, the channel estimation error will be added to the existing receiver impairments, thus increasing the residual self-interference power. Figure~\ref{Fig5Label}(a) shows the increment in the residual self-interference power due to imperfect channel estimation as compared to the case of perfect channel knowledge. The results show that, the residual self-interference power is doubled when the channel is estimated using one training symbol. However, averaging over 4 symbols reduces the degradation to only 1dB compared to the perfect channel case

Another important factor is the channel estimation error due to the time varying nature of the wireless channel. In practical indoor applications, the channel is estimated once at the beginning of each data frame, and then assumed to be constant within the frame. While this might be sufficient for half-duplex systems, this assumption does not hold for the self-interference channel in full-duplex systems. The reason is that, in half-duplex systems, the signal-of-interest arrives at relatively small power levels such that the channel error due to fading effect will be much smaller than the noise floor. However, in full-duplex systems, the power of the self-interference signal is significantly high, such that the channel error due to the fading effect might dominate other noise components, especially with long data frames.

Figure~\ref{Fig5Label}(b) show the increment in the residual self-interference power due to the fading effect as compared to the case of perfect channel knowledge. The fading effect is investigated at different frame lengths with 5Hz Doppler frequency channel. The results show that for long data frames, the fading effect significantly increases the residual self-interference power compared to the perfect channel case. More specifically, up to 6.5dB performance degradation is expected for frame lengths of 150 symbols (i.e. 600us based on the simulated system parameters). It can also be noticed that the performance degradation is decreasing with the decrease of the received signal power level. The reason is that, at low received signal power levels, the fading effect becomes negligible compared to the receiver impairments and the performance becomes noise limited. However, at high received signal power levels, the fading effect dominates other noise components. This behavior also explains why the fading effect is a significant concern in full-duplex systems, while it is not a concern in half-duplex systems.

From figures~\ref{Fig5Label}(a),~(b) we can conclude that reducing the channel estimation error requires more training symbols and shorter frame lengths. Both requirements negatively impact the overall system capacity. In fact, capacity degradation is related to the training overhead, which is defined as the ratio between the number of training symbols to the number of useful data symbols. At the same training overhead, several combinations of the number of training symbols and frame lengths could be used. For example, 4\% training overhead could be achieved with 2 training symbols every 50 data symbols, or 4 training symbols every 100 data symbols, etc. The appropriate choice should be made based on the receiver operating point, mainly the received signal power level. For instance, at high received signal power levels, using shorter frames with a small number of training symbols is better than using long frames with a large number of training symbols and vice versa. This tradeoff is clear in figures~\ref{Fig5Label}(a),~(b), where it is shown that at high received signal power levels, doubling the frame length from 50 to 100 symbols degrades the performance by 2.5dB, while reducing the number of training symbols from 4 to 2 symbols only degrades the performance by 1dB. In this scenario, for 4\% training overhead, using 50-symbols frame with 2 training symbols is better than using 100-symbols frame with 4 training symbols. On the other hand, at low received signal power levels, the degradation due to the fading effect is smaller than the degradation due to the receiver impairments. In this case, using more training symbols is more beneficial than using short frames. 
\subsection{Impact of Receiver Nonlinearities}
In the presence of only receiver nonlinearities, the auxiliary and ordinary receiver outputs (Equation~\eqref{eq:2}~and~\eqref{eq:3}) can be rewritten as
\begin{equation}\label{eq:30}
y_n^{aux}=y_n^{tx}*h_n^{aux}\text{.}
\end{equation}
\begin{equation}\label{eq:31}
y_n^{ord}=y_n^{tx}*h_n^{ord}+d_n^{rx}+s^{soi}_n\text{.}
\end{equation}
Then, after self-interference cancellation, the interference-free signal $Y^{DC}$ can be written as
\begin{equation}\label{eq:32}
Y_k^{DC}=Y_k^{ord}-\frac{H_k^{ord}}{H_k^{aux}}Y_k^{aux}=Y_k^{tx} H_k^{ord}+D_k^{rx}+S_k^{soi}-\frac{H_k^{ord}}{H_k^{aux}}\left(Y_k^{tx} H_k^{aux}\right)= D_k^{rx}+S_k^{soi}\text{,}
\end{equation}
Where $D^{rx}_k$  is the frequency domain representation of the LNA nonlinear distortion signal $d^{rx}$. In this analysis, we are assuming that the LNA is the major contributor to the receiver nonlinearity. Therefore, the auxiliary receiver chain is assumed to be highly linear, since there is no LNA in the auxiliary receiver chain. According to~\eqref{eq:32}, it can be easily shown that the residual self-interference power is directly proportional to the nonlinear distortion level. Accordingly, receiver nonlinearity suppression is essential for better self-interference cancellation capability.

One technique for nonlinearity estimation and suppression in full-duplex systems is proposed by the authors in~\cite{Ref8}. In this technique, a LS estimator is used to jointly estimate the transmitter and receiver nonlinearity coefficients. Then, based on the estimated coefficients, the nonlinear distortion signal is reconstructed and subtracted from the received signal. The technique presented in~\cite{Ref8} could be modified to be used with the proposed architecture for receiver nonlinearity suppression as follows. 

According to the nonlinearity model in~\eqref{eq:7}, and considering only third order nonlinearities, the receiver nonlinear distortion signal $d^{rx}$ can be written in terms of the transmitted signal $y^{tx}$ as 
\begin{equation}\label{eq:33}
d_n^{rx}=\alpha_3\left(y_n^{tx}*h_n^{ord}\right)\left|y_n^{tx}*h_n^{ord}\right|^2\text{.}
\end{equation}
Using~\eqref{eq:1}, the base-band representation of the transmitted signal $y^{tx}(t)$ can be written as
\begin{equation}\label{eq:34}
y_n^{tx}=x_ne^{j\phi_n^{tx}}+d_n^{tx}=x_n+jx_n\phi_n^{tx}+d_n^{tx}\text{,}
\end{equation}
Where $e^{j\phi} \simeq 1+j\phi ,  \phi \ll 1$. The second and third terms in the right hand side of (34) is mainly the transmitter phase noise and distortion signals that are typically $\ll x_n$ in terms of power. According to~\eqref{eq:34}, the transmitter phase noise and distortion signals will contribute to the receiver nonlinear distortion $d^{rx}$. However, their contribution will be much smaller than the distortion due to the main transmitted signal $x_n$. Therefore, the transmitter phase noise and distortion signals could be ignored while substituting from~\eqref{eq:34} into~\eqref{eq:33}. Accordingly, equation~\eqref{eq:33} can be approximated as
\begin{equation}\label{eq:35}
d_n^{rx}=\alpha_3\left(x_n*h_n^{ord}\right)\left|x_n*h_n^{ord}\right|^2\text{.}
\end{equation}
In~\eqref{eq:35}, $x_n$ is known and $h^{ord}$ could be estimated during the channel estimation training period. The only unknown in~\eqref{eq:35} is the nonlinearity coefficient $\alpha_3$, which could be estimated using simple LS estimator. As described before, channel estimation error is directly proportional to the receiver impairments. Accordingly, in the presence of receiver nonlinearity, the channel estimation error will be also a function of the receiver nonlinearity. Therefore, any nonlinearity estimation and suppression technique should consider the channel estimation error due to receiver nonlinearity.

Referring back to~\eqref{eq:30} and~\eqref{eq:31}, the channel estimation in the presence of only receiver nonlinearity is performed as
\begin{equation}\label{eq:36}
\hat{H}_x=\frac{Y_k^{ord}}{Y_k^{aux}}=\frac{Y_k^{tx}H_k^{ord}+D_k^{rx}}{Y_k^{tx}H_k^{aux}}=\frac{H_k^{ord}}{H_k^{aux}}+\frac{D_k^{rx}}{Y_k^{tx}H_k^{aux}}\text{.}
\end{equation}
In order to estimate the nonlinearity coefficient, one additional training symbol is transmitted after the channel estimation training symbol. Superscript $tr1$, and $tr2$ is given to the different signals within the channel estimation training symbol and the additional training symbol respectively. During the additional training symbol, the received signal after self-interference cancellation can be written as
\begin{eqnarray}\label{eq:37}
Y_k^{DC,tr2} &=& Y_k^{ord,tr2}-\hat{H}_k Y_k^{aux,tr2} = Y_k^{tx,tr2}H_k^{ord}+D_k^{rx,tr2}-\left(\frac{H_k^{ord}}{H_k^{aux}}+\frac{D_k^{rx,tr1}}{Y_k^{tx,tr1}H_k^{aux}}\right)\left(Y_k^{tx,tr2}H_k^{aux}\right) \nonumber \\
&=& D_k^{rx,tr2}-\frac{Y_k^{tx,tr2}}{Y_k^{tx,tr1}}D_k^{rx,tr1} \text{.}
\end{eqnarray}
Using the same approximation as in~\eqref{eq:35}, equation~\eqref{eq:37} can be approximated as
\begin{equation}\label{eq:38}
Y_k^{DC,tr2}=D_k^{rx,tr2}-\frac{X_k^{tr2}}{X_k^{tr1}}D_k^{rx,tr1}=\alpha_3\left(\bar{D}_k^{rx,tr2}-\frac{X_k^{tr2}}{X_k^{tr1}}\bar{D}_k^{rx,tr1}\right) \text{,}
\end{equation}
and
\begin{equation}\label{eq:39}
\bar{D}_k^{rx,i} = DFT\left[\left(x_n^i*h_n^{ord}\right)\left|x_n^i*h_n^{ord}\right|^2\right], \ \ \ i\in\{tr1,tr2\} \text{.}
\end{equation}
Finally, an estimate for the nonlinearity coefficient $\alpha_3$ is obtained as
\begin{equation}\label{eq:40}
\hat{\alpha}_3 = \frac{1}{N}\sum_{k=0}^{N-1}\frac{Y_k^{DC,tr2}}{\bar{D}_k^{rx,tr2}-\frac{X_k^{tr2}}{X_k^{tr1}}\bar{D}_k^{rx,tr1}}\text{.}
\end{equation}

During the data symbols, the estimated nonlinearity coefficient $\hat{\alpha}_3$ is used to reconstruct the distortion signal and subtract it from the received signal. Note that, the nonlinearity coefficients change very slowly such that it could be estimated once every several frames, which reduces the training overhead due to the use of the additional training symbol.

\begin{figure}[!ht]
\begin{center}
\noindent
  \includegraphics[width=3.5in]{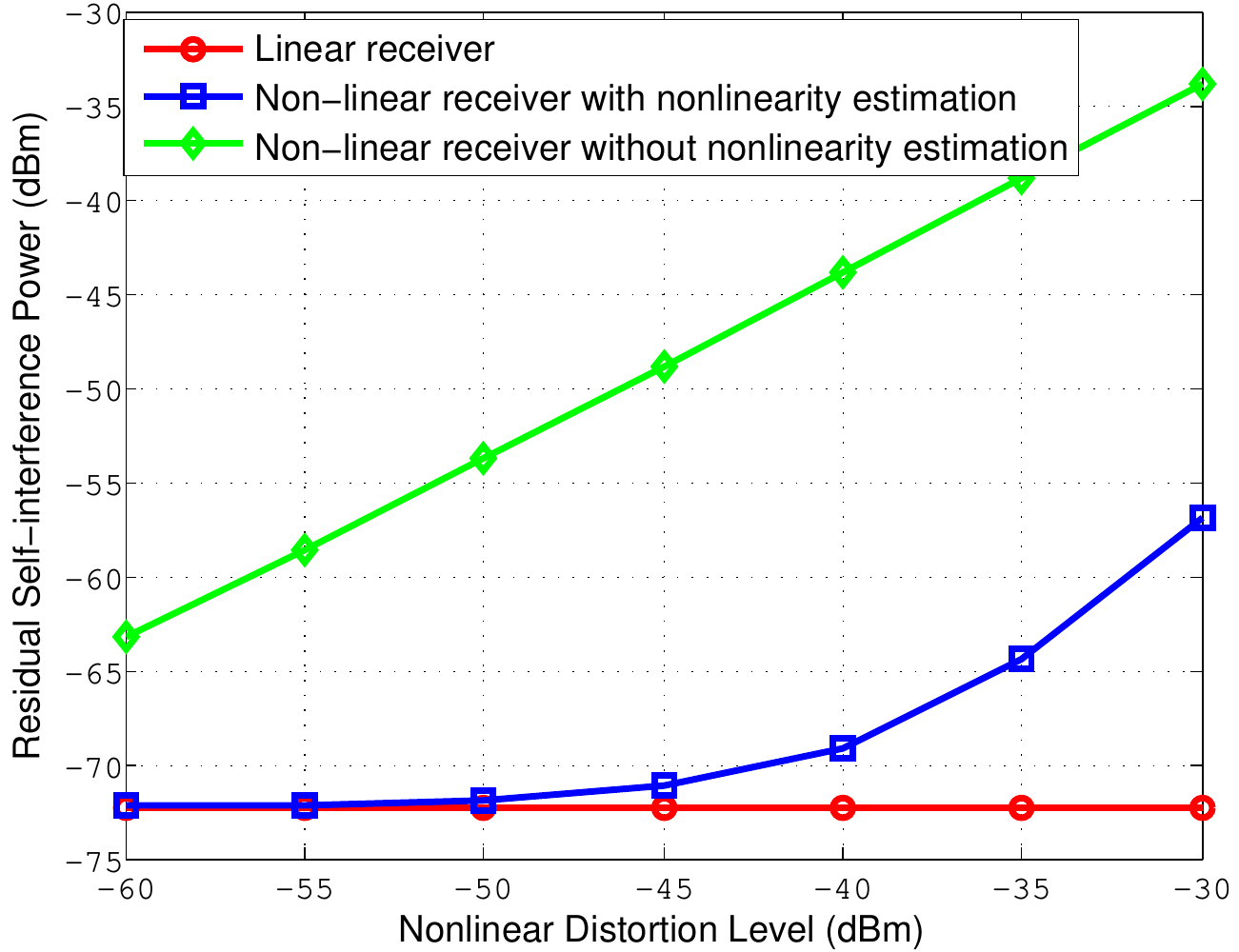}
  \caption{Residual self-interference power due to receiver nonlinearities with and without nonlinearity estimation.\label{Fig6Label}}
\end{center}
\end{figure}

The proposed nonlinearity estimation and suppression technique is numerically investigated at different operating conditions. Figure~\ref{Fig6Label} shows the residual self-interference power at different nonlinear distortion power levels in three cases: i) the linear receiver case, ii) the nonlinear receiver case without performing nonlinearity suppression, and iii) the nonlinear receiver case with the proposed nonlinearity suppression technique. In the linear receiver case, the performance is limited by other receiver impairments and channel estimation errors. The results show that without nonlinearity suppression, the residual self-interference power is limited by the distortion power level. However, performing nonlinearity suppression using the proposed technique could achieve $\sim$23dB more reduction in the residual self-interference power. Furthermore, at distortion power levels $\leq -$45dBm, the proposed technique is shown to achieve almost the same performance as in the linear receiver case, which means that the 23dB improvement are sufficient to eliminate the nonlinearity effect at such distortion power levels. 
\subsection{Overall Cancellation Performance}
In this section, the overall self-interference cancellation performance of the proposed technique is numerically investigated in the presence of all transmitter and receiver impairments. The design tradeoffs and the factors contributing to the residual self-interference power are discussed. In the previous analyses we show that the residual self-interference power highly depends on the following factors: i) the power level of each one of the receiver impairments, ii) the received self-interference power level, iii) the channel Rician factor which determines the power of the received non-LOS component, and iv) the self-interference channel characteristics (e.g. coherence bandwidth, Doppler frequency). In practical full-duplex systems, those four parameters are not totally independent. For instance, the experimental results in~\cite{Ref13} show that, the self-interference channel Rician factor is inversely proportional to the achieved passive suppression amount which determines the received self-interference signal power. Accordingly, for reliable conclusions, such dependency should be considered while investigating the overall cancellation performance.

To be pragmatic, all presented system parameters are chosen based on practical transceivers and real-time experimental results. More specifically, the values for the transmitter and receiver impairments are chosen based on the datasheet of the NI5791 transceiver~\cite{Ref25} as follows: i) the integrated in-band transmitter and receiver phase noise power is -40dBc, ii) the number of ADC bits is 14bits, iii) the Gaussian noise power is $-$90dBm and $-$72dBm at receiver input power levels of $-$25dBm and $-$5dBm respectively, and iv) the transmitter and receiver third order distortion power level is $-$45dB from the linear component power level. The system is simulated with the indoor TGn channel model 'D'~\cite{Ref26} at 5Hz Doppler frequency. The training overhead is set to 4\%.

From passive suppression perspective, three practical scenarios are investigated: a) the use of omni-directional antenna with 35cm antenna separation. In this scenario, 25dB of passive self-interference suppression is achieved, and the self-interference channel Rician factor is 20dB~\cite{Ref13}. b) The use of directional antennas with absorbing material between the transmit and receive antennas~\cite{Ref13}. In this secnario, 45dB of passive suppression is achieved, and the self-interference channel Rician factor drops to 0dB. c) The use of reconfigurable directional antennas~\cite{Ref14}. In this secnario, up to 60dB of passive suppression is achieved, and the self-interference channel Rician factor is 0dB. The main difference between the three scenarios is the received self-interference power, and the received non-LOS component power level. For example, at transmit power of 20dBm, the first scenario will have a received self-interference power of $-$5dBm and non-LOS component power level of $-$25dBm. While in the second scenario, both the received self-interference power and the non-LOS component power levels are at $-$25dBm.

\begin{figure}[!ht]
\begin{center}
\noindent
  \includegraphics[width=6.5in]{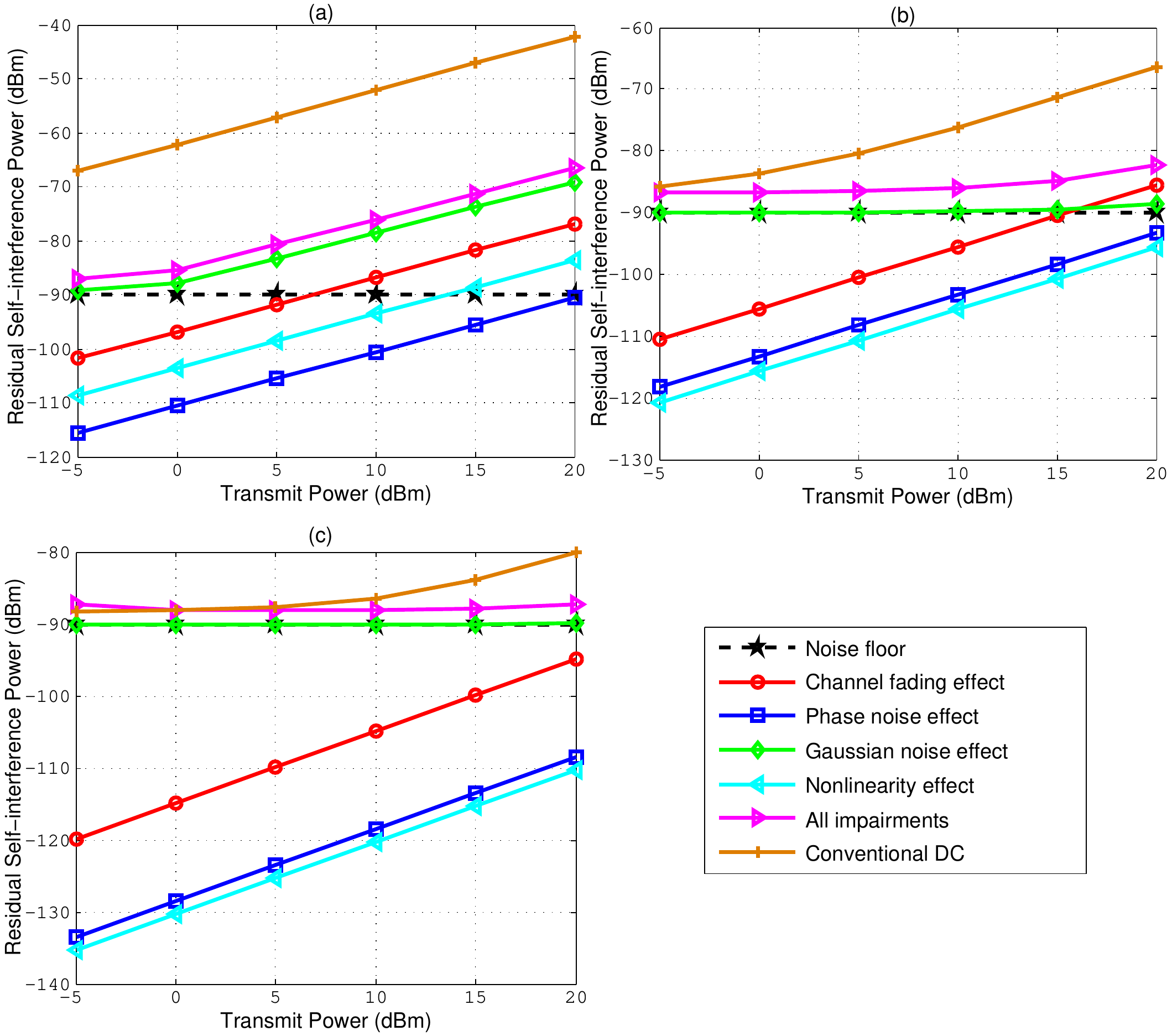}
  \caption{Residual self-interference power due to different receiver impairments at different transmit power values: (a) $1^{st}$ scenario, (b) $2^{nd}$ scenario, (c) $3^{rd}$ scenario.\label{Fig7Label}}
\end{center}
\end{figure}

In each scenario, the performance of the proposed digital cancellation technique is investigated at different transmit power values. The performance is also compared to the conventional digital cancellation techniques~\cite{Ref6}. The residual self-interference power due to each one of the receiver impairments is also presented to identify the main bottleneck in each region of operation. In all analyses, the half-duplex system's noise floor is shown for comparison purposes. Figure~\ref{Fig7Label}(a)-(c) shows the residual self-interference power at different transmit power values for the three scenarios. The conclusions from these results are multifold: first, in all scenarios, the proposed self-interference cancellation technique significantly mitigates the phase noise and nonlinearity effects to below the receiver noise floor.  Therefore, in contrast with conventional digital cancellation techniques, the cancellation capability of the proposed techniques is no longer phase noise or nonlinearity limited.

Second, in the first scenario, due to the relatively high received self-interference power, the receiver Gaussian noise dominates other noise components and becomes the performance limiting factor. However, as the transmit power decreases, the Gaussian noise decreases which reduces the residual self-interference power. Accordingly, in such scenarios with relatively low passive suppression amounts, the proposed technique is more suitable to be used in low transmit power applications (e.g. up to 5dBm transmit power levels). Furthermore, simple analog cancellation techniques (e.g. Balun technique~\cite{Ref1}) could be used to alleviate the Gaussian noise effect in such scenarios. On the other hand, when good passive suppression techniques are used (e.g. second and third scenarios), the Gaussian noise is no longer the limiting factor, and the self-interference signal could be significantly mitigated to $\sim$3dB above the half-duplex system's noise floor. 

Third, following the receiver Gaussian noise, the channel error due to the fading effect is found to be the next performance bottleneck. The good thing about the error due to the fading effect is that, in contrast to other receiver impairments, there are many ways to reduce the fading effect. For example, i) interpolating the channel between different data frames, ii) inserting pilots within the OFDM symbol to track the channel variations, iii) using shorter frames lengths, or iv) using pilot-based frame structure instead of the preamble-based frame structure. In the pilot-based frame structures, pilot subcarriers are inserted within the OFDM symbols to be used for the channel estimation purposes. Such pilot subcarriers allow for fast tracking of the channel variations. 

\section{Achievable Rate Analysis}
In this section, the overall full-duplex system performance using the proposed digital cancellation technique is investigated in terms of the achievable rate gain compared to conventional half-duplex systems. The performance is investigated in the same three operating scenarios described in Section III-E. Both full-duplex and half-duplex system performances are investigated in the presence of all transmitter and receiver impairments. The performance is investigated at two transmit power values: 20dBm, and 5dBm. 

Generally, the achievable rate for both full-duplex and half-duplex systems can calculated as 
\begin{equation}\label{eq:41}
R^{FD}=E\left[log_2\left(1+SINR^{FD}\right)\right]\text{,}
\end{equation}
\begin{equation}\label{eq:42}
R^{HD}=\frac{1}{2} E\left[log_2\left(1+SNR^{HD}\right)\right]\text{,}
\end{equation}
where $E[\ ]$ denotes expectation process, $R^{FD}$, $R^{HD}$ is the full-duplex and half-duplex system's achievable rate, $SINR^{FD}$ is the signal-of-interest to interference plus noise ratio in full-duplex systems, and $SNR^{HD}$ is the signal to noise ratio in half-duplex systems. The factor of $1/2$ in the half-duplex rate equation is due to the fact that the resources are divided between the two communicating nodes. 

\begin{figure}[!ht]
\begin{center}
\noindent
  \includegraphics[width=6.5in]{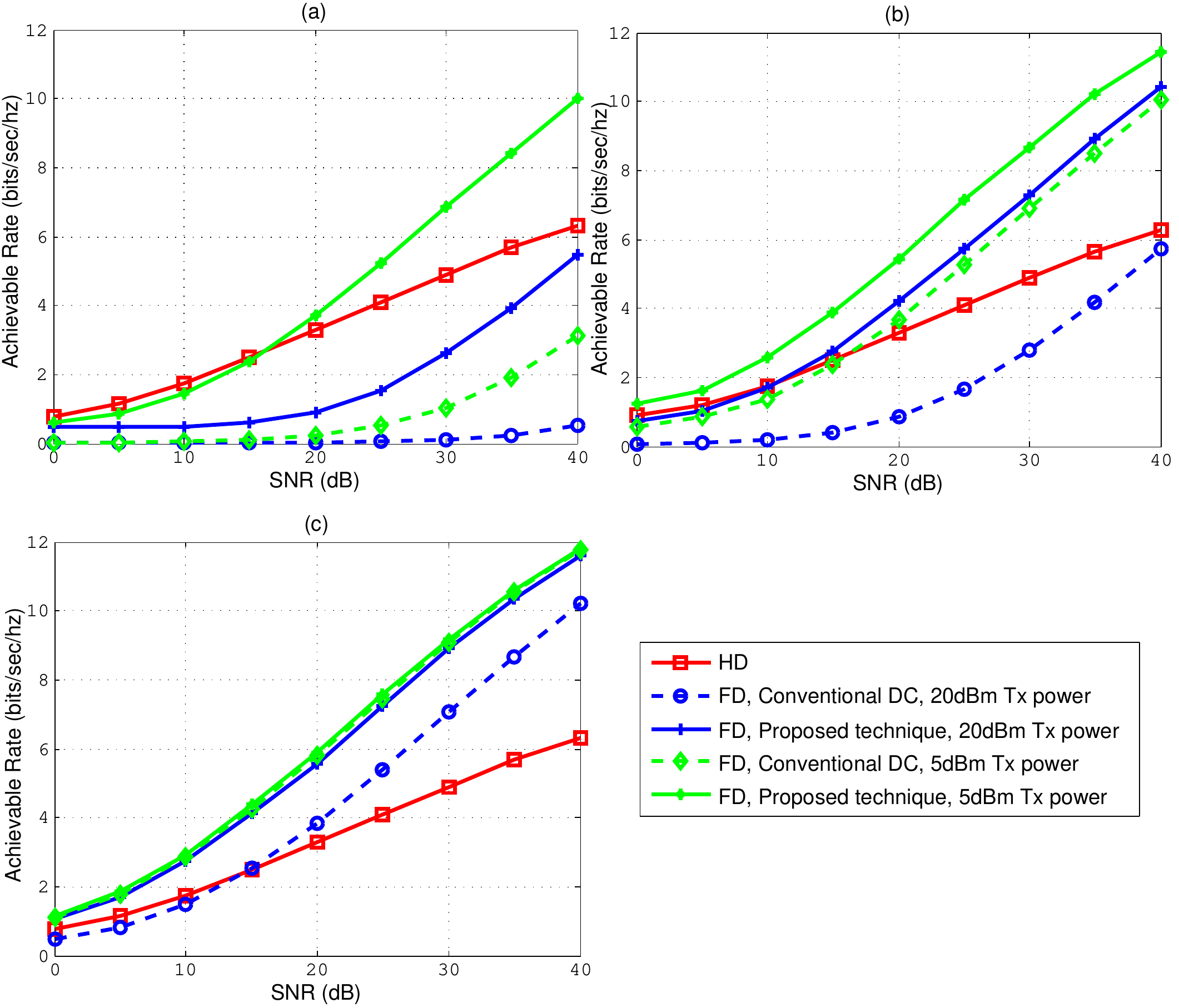}
  \caption{Full-duplex and Half-duplex acheivable rates at different transmit power and SNR values: (a) $1^{st}$ scenario, (b) $2^{nd}$ scenario, (c) $3^{rd}$ scenario.\label{Fig8Label}}
\end{center}
\end{figure}

Figure~\ref{Fig8Label}(a)-(c) shows the achievable rate for the three scenarios at different SNR values. The conclusions from these results are multifold: first, the results show that the proposed technique significantly outperforms the conventional digital cancellation technique in all operating scenarios. The only exception is the case of 5dBm transmit power in the third scenario, where both the proposed technique and the conventional digital cancellation technique achieves the same performance. The reason is that, at 5dBm transmit power with the 60dB passive suppression assumed in the third scenario, only 35dB of digital cancellation are required to suppress the self-interference signal to the noise floor. Those 35dB could be easily achieved using conventional digital cancellation techniques. However, as the transmit power increases, the proposed technique will be able to achieve more self-interference cancellation, while the cancellation achieved using the conventional techniques will saturate. Second, figure~\ref{Fig8Label}(a) shows that, in the first scenario where the passive suppression amount is relatively low, the proposed technique can only be used in low transmit power applications.

Third, in the second and third scenarios, the proposed technique achieves significant rate improvement compared to conventional half-duplex systems, especially in high SNR regimes. In order to quantify the achievable rate gain of the proposed full-duplex system compared to the conventional half-duplex system, we calculate the average rate gain over the whole SNR range from 0 to 40dB for all scenarios, and the results are shown in table~\ref{Table1Label}. The results in table~\ref{Table1Label} are average results; however, exact rate gain for each SNR value could be obtained from figure~\ref{Fig8Label}(a)-(c). Negative rate gain means that operating in full-duplex mode will degrade the overall system performance.

\section{Conclusion}
In this paper, a novel all-digital self-interference cancellation technique for full-duplex systems is proposed. The proposed technique uses an auxiliary receiver chain to obtain a digital-domain copy of the transmitted RF self-interference signal including all transmitter impairments. The self-interference signal copy is then used in the digital-domain to cancel out both the self-interference signal and the transmitter impairments. In order to alleviate the receiver phase noise effect, a common oscillator is shared between the auxiliary and ordinary receiver chains. In addition, a nonlinearity estimation and suppression technique is proposed to mitigate the receiver nonlinearity effects. A thorough analytical and numerical analysis for the effect of the transmitter and receiver impairments on the cancellation capability of the proposed technique is presented. The analyses show that the proposed technique significantly mitigates the transceiver phase noise and nonlinearity effects. The overall full-duplex system performance using a combination of the proposed digital cancellation technique and practical passive suppression techniques is numerically investigated. The results show that, the proposed technique significantly mitigates the self-interference signal to $\sim$3dB higher than the receiver noise floor, which results in up to 76\% rate improvement compared to conventional half-duplex systems at 20dBm transmit power values.
\clearpage 
\begin{table}[ht]
\caption{Average full-duplex rate improvement compared to half-duplex system in different operating scenarios}
\label{Table1Label}
\centering
\begin{tabular}{|c|c|c|c|c|c|c|}
\hline
 & \multicolumn{2}{c|}{$1^{st}$ Scenario} & \multicolumn{2}{c|}{$2^{nd}$ Scenario} & \multicolumn{2}{c|}{$3^{rd}$ Scenario} \\
& \multicolumn{2}{c|}{(25dB passive suppression)} & \multicolumn{2}{c|}{(45dB passive suppression)} & \multicolumn{2}{c|}{(60dB passive suppression)} \\
 \hline
 & Proposed & Conventional & Proposed & Conventional & Proposed & Conventional \\
& Technique & DC & Technique & DC & Technique & DC \\
 \hline
 5dBm Transmit Power& 14\%  &   $-$58\%  &  61\%  & 11\%  &  76\%  & 72\% \\ 
\hline
 20dBm Transmit Power&  $-$58\% & $-$98\% & 23\% &  $-$63\% &  67\% & 14\%  \\ 
\hline
\end{tabular}
\end{table}
%



\begin{thebibliography}{1}

\bibitem{Ref1}
M. Jain, J. I. Choi, T. Kim, D. Bharadia, K. Srinivasan, S. Seth, P. Levis, S. Katti, and P. Sinha, ``Practical, Real-time, Full Duplex Wireless," \emph{in Proceeding of the ACM Mobicom}, Sept. 2011.

\bibitem{Ref2}
B. Radunovic, D. Gunawardena, P. Key, A. P. N. Singh, V. Balan, and G. Dejean, ``Rethinking indoor wireless Mesh Design: Low power, low frequency, full duplex," \emph{Wireless Mesh Networks (WIMESH 2010), 2010 Fifth IEEE Workshop on}, pp.1-6, June 2010.

\bibitem{Ref3}
M.E. Knox, ``Single antenna full duplex communications using a common carrier," \emph{Wireless and Microwave Technology Conference (WAMICON), 2012 IEEE 13th Annual}, 15-17 April 2012.

\bibitem{Ref4}
A. Balatsoukas-Stimming, P. Belanovic, K. Alexandris, and A. Burg, ``On Self-interference Suppression Methods for Low-complexity Full-duplex MIMO," \emph{Signals, Systems and Computers, 2013 Asilomar Conference on}, pp.992,997, Nov. 2013.

\bibitem{Ref5}
M. Duarte, and A. Sabharwal, ``Full-duplex wireless communications using off-the-shelf radios: Feasibility and first results," \emph{Signals, Systems and Computers, 2010 Asilomar Conference on}, pp.1558-1562, Nov. 2010.

\bibitem{Ref6}
M. Duarte, C. Dick, and A. Sabharwal, ``Experiment-Driven Characterization of Full-Duplex Wireless Systems," \emph{Wireless Communications, IEEE Transactions on}, vol.11, no.12, pp.4296,4307, December 2012.

\bibitem{Ref7}
D. Bharadia, E. McMilin, and S. Katti, ``Full Duplex Radios, " in \emph{ACM SIGCOMM}, Aug. 2013.

\bibitem{Ref8}
E. Ahmed, A. M. Eltawil, and A. Sabharwal, ``Self-Interference Cancellation with Nonlinear Distortion Suppression for Full-Duplex Systems," \emph{Signals, Systems and Computers, 2013 Asilomar Conference on}, pp.1199,1203, Nov. 2013.

\bibitem{Ref9}
E. Ahmed, A. M. Eltawil, and A. Sabharwal, ``Self-Interference Cancellation with Phase Noise Induced ICI Suppression for Full-Duplex Systems," \emph{Global Communications Conference (GLOBECOM), 2013 IEEE}, pp.3384,3388, Dec. 2013.

\bibitem{Ref10}
J. I. Choi, M. Jain, K. Srinivasan, P. Levis, and S. Katti, ``Achieving single channel, full duplex wireless communication," \emph{in MobiCom}, 2010.

\bibitem{Ref11}
E. Everett, M. Duarte, C. Dick, and A. Sabharwal, ``Empowering full-duplex wireless communication by exploiting directional diversity," \emph{Signals, Systems and Computers, 2011 Asilomar Conference on}, pp.2002-2006, Nov. 2011.

\bibitem{Ref12}
E. Ahmed, A. M. Eltawil, and A. Sabharwal, ``Simultaneous transmit and sense for cognitive radios using full-duplex: A first study," \emph{Antennas and Propagation Society International Symposium (APSURSI), 2012 IEEE} , pp.1-2, July 2012.

\bibitem{Ref13}
E. Everett, A. Sahai, and A. Sabharwal, ``Passive Self-Interference Suppression for Full-Duplex Infrastructure Nodes," \emph{Wireless Communications, IEEE Transactions on}, vol.13, no.2, pp.680,694, February 2014.

\bibitem{Ref14}
E. Ahmed, A. M. Eltawil, Z. Li, and B. A. Cetiner ``Full-Duplex Systems Using Multi-Reconfigurable Antennas," submitted to \emph{IEEE Transactions on Wireless Communications}. [Online]. Available: http://arxiv.org/abs/1405.7720.

\bibitem{Ref15}
M. Duarte, A. Sabharwal, V. Aggarwal, R. Jana, K. Ramakrishnan, C. Rice, and N. Shankaranayanan, ``Design and Characterization of a Full-Duplex Multiantenna System for WiFi Networks," \emph{Vehicular Technology, IEEE Transactions on}, vol.63, no.3, pp.1160,1177, March 2014.

\bibitem{Ref16}
E. Ahmed, A. M. Eltawil, and A. Sabharwal, ``Rate Gain Region and Design Tradeoffs for Full-Duplex Wireless Communications, " \emph{Wireless Communications, IEEE Transactions on} , vol.12, no.7, pp.3556,3565, July 2013.

\bibitem{Ref17}
A. Sahai, G. Patel, C. Dick, and A. Sabharwal, ``On the Impact of Phase Noise on Active Cancelation in Wireless Full-Duplex," \emph{Vehicular Technology, IEEE Transactions on}, vol.62, no.9, pp.4494,4510, Nov. 2013.

\bibitem{Ref18}
D.W. Bliss, T.M. Hancock, and P. Schniter, ``Hardware phenomenological effects on cochannel full-duplex MIMO relay performance," \emph{Signals, Systems and Computers, 2012 Asilomar Conference on}, pp.34,39, 4-7 Nov. 2012.

\bibitem{Ref19}
E. Ahmed, and A. M. Eltawil, ``On Phase Noise Suppression in Full-Duplex Systems," submitted to \emph{IEEE Transactions on Wireless Communications}. [Online]. Available: http://arxiv.org/abs/1401.6437.

\bibitem{Ref20}
T. Schenk, RF Imperfections in High-rate Wireless Systems, Impact and Digital Compensation. New York: Springer-Verlag, 2008.

\bibitem{Ref21}
B. Razavi, Design of Analog CMOS Integrated Circuits. New York:McGraw-Hill, 2001.

\bibitem{Ref22}
A. Demir, A. Mehrotra, and J. Roychowdhury, ``Phase noise in oscillators; a unifying theory and numerical methods for characterization," \emph{IEEE Trans. Circuits Syst. I}, vol. 47, pp. 655-674 May 2000.

\bibitem{Ref23}
A. Mehrotra, ``Noise analysis of phase-locked loops," \emph{IEEE Trans. Circuits Syst. I}, vol. 49,no. 9, pp. 1309-1316 Sep. 2002.

\bibitem{Ref24}
M. Lu, N. Shanbhag, and A. Singer, ``BER-optimal analog-to-digital converters for communication links," \emph{Circuits and Systems (ISCAS), Proceedings of 2010 IEEE International Symposium on}, pp.1029-1032, May 30 2010-June 2 2010.

\bibitem{Ref25}
NI 5791R User Manual and Specifications. http://www.ni.com/pdf/manuals/373845d.pdf.

\bibitem{Ref26}
V. Erceg, and L. Schumacher et al., ``TGn channel models," IEEE 802.11-03/940r4, May 2004.

\bibitem{Ref27}
``IEEE Standard for Information technology--Telecommunications and information exchange between systems Local and metropolitan area networks--Specific requirements Part 11: Wireless LAN Medium Access Control (MAC) and Physical Layer (PHY) Specifications," \emph{IEEE Std 802.11-2012 (Revision of IEEE Std 802.11-2007)}, March 2012.

\end{thebibliography}
\end{document}